\documentclass[conference]{IEEEtran}
\IEEEoverridecommandlockouts
\usepackage{cite}
\usepackage{amsmath,amssymb,amsfonts}
\usepackage{algorithmic}
\usepackage{graphicx}
\usepackage{textcomp}
\usepackage{xcolor}
\def\BibTeX{{\rm B\kern-.05em{\sc i\kern-.025em b}\kern-.08em
    T\kern-.1667em\lower.7ex\hbox{E}\kern-.125emX}}

\usepackage{bm}
\usepackage{subfigure}
\usepackage{multirow}
\usepackage{array}
\newcolumntype{C}[1]{>{\centering\let\newline\\\arraybackslash\hspace{0pt}}m{#1}}

\begin{document}

\title{LINN: Lifting Inspired Invertible Neural Network for Image Denoising\\
}

\author{\IEEEauthorblockN{Jun-Jie Huang}
\IEEEauthorblockA{\textit{Electrical and Electronic Engineering Department} \\
\textit{Imperial College London}\\
London, U.K. \\
j.huang15@imperial.ac.uk}
\and
\IEEEauthorblockN{Pier Luigi Dragotti}
\IEEEauthorblockA{\textit{Electrical and Electronic Engineering Department} \\
\textit{Imperial College London}\\
London, U.K. \\
p.dragotti@imperial.ac.uk}
}

\maketitle

\begin{abstract}
In this paper, we propose an invertible neural network for image denoising (DnINN) inspired by the transform-based denoising framework. The proposed DnINN consists of an invertible neural network called LINN whose architecture is inspired by the lifting scheme in wavelet theory and a sparsity-driven denoising network which is used to remove noise from the transform coefficients. 
The denoising operation is performed with a single soft-thresholding operation or with a learned iterative shrinkage thresholding network. The forward pass of LINN produces an over-complete representation which is more suitable for denoising. The denoised image is reconstructed using the backward pass of LINN using the output of the denoising network.
The simulation results show that the proposed DnINN method achieves results comparable to the DnCNN method while only requiring 1/4 of learnable parameters.

\end{abstract}


\section{Introduction}

Image denoising is a classical problem in signal processing and computer vision. The objective is to restore an unknown noiseless image $\bm{x} \in \mathbb{R}^{m \times n}$ from the observed noisy image $\bm{y} \in \mathbb{R}^{m \times n}$ which can be modeled as:
\begin{equation}
    \bm{y} = \bm{x} + \bm{e},
\end{equation}
where $\bm{e} \in \mathbb{R}^{m \times n}$ is the additive white Gaussian noise.

In the past decades, many algorithms have been proposed for image denoising, including wavelet-domain denoising \cite{donoho1994ideal, donoho1995adapting, chang2000adaptive, blu2007sure}, non-local means \cite{buades2005non, mahmoudi2005fast}, sparse dictionaries \cite{elad2006image, dong2011sparsity}, BM3D \cite{dabov2007image}, weighted nuclear norm minimization (WNNM) \cite{gu2014weighted} and deep neural networks (DNNs) \cite{burger2012image, chen2016trainable, zhang2017beyond, zhang2018ffdnet}. The classical image denoising algorithms \cite{donoho1994ideal, donoho1995adapting, chang2000adaptive, blu2007sure, buades2005non, mahmoudi2005fast, elad2006image, dong2011sparsity, dabov2007image, gu2014weighted} have good denoising performance and solid theoretical supports. The recent DNNs-based methods \cite{burger2012image, chen2016trainable, zhang2017beyond, zhang2018ffdnet} achieve state-of-the-art results but are not easy to interpret.

Redundant transforms and sparsity have been shown to be a simple yet powerful tool for image denoising. The sparse representation theory also provides a good way to interpret the algorithm. Within this context, the noisy image is first transformed and then non-linear thresholding is applied on the transform coefficients. The denoised image is then obtained through performing inverse transform on the thresholded coefficients. Both wavelet-domain \cite{donoho1994ideal, donoho1995adapting, blu2007sure} and sparse dictionaries \cite{elad2006image, dong2011sparsity} based image denoising algorithms assume a linear transform which may limit the image denoising performance. Therefore, exploring a redundant non-linear transform and the connection with sparsity could have the potential to lead to simple image denoising algorithms with good performances as well as a good interpretability.

The lifting scheme introduced in \cite{sweldens1998lifting, daubechies1998factoring} proposes to construct a wavelet transform by first splitting the signal into an even and an odd part. A predictor is used to predict the odd signal from the even part, therefore leading to a sparse signal. The update step is used to adjust the even signal based on the prediction error of the odd part to make it a better coarse version of the original signal. There can be multiple pairs of predictor and updater. Though the prediction and the update can be arbitrary complex function, the lifting scheme can represent a transform with perfect reconstruction condition. Any intermediate representation can be used to infer the input signal and the final representation.
Inspired by this scheme, the invertible neural networks (INNs) \cite{dinh2014nice, dinh2016density, jacobsen2018revnet} was proposed for memory-efficient backpropagation model and used for constructing the flow-based generative models.

In this paper, we propose an image denoising invertible neural network (DnINN) based on the principle of transform-based denoising. It consists of a lifting inspired invertible neural network (LINN) and a sparsity-driven denoising network. The LINN is used to construct a non-linear transform which can well separate the image content from the noise. The forward pass of a lifting inspired invertible neural network generates a redundant representation of the input noisy image. The simple denoising network aims to denoise the transform coefficients by removing the features corresponding to noise. The denoised image is then constructed using the backward pass of the lifting inspired invertible neural networks. 

Since invertible neural networks construct a bijective function, the representation produced by INN contains all information of its input. In the proposed DnINN, the only non-invertible part is the denoising network, therefore, we can easily interpret the workings and it is possible to adapt the learned DnINN to different noise levels by adjusting the thresholds in the denoising network.

The rest of the paper is organized as follows: Section \ref{sec:method} introduces the implementation details of our proposed image denoising invertible neural network, Section \ref{sec:results} shows and analyzes the numerical results of the proposed method and finally Section \ref{sec:conclude} draws the conclusions.

\section{Proposed Method} \label{sec:method}

In this section, we will first provide an overview of the proposed image denoising invertible neural network (DnINN), we will then introduce the implementation details of the network, and the training details.

\subsection{Overview}\label{sec:overview}

The proposed DnINN method comprises a lifting inspired invertible neural network (LINN) and a sparsity-driven denoising network. LINN implements a non-linear transform with perfect reconstruction property and the denoising network implements sparse coding. Both LINN and the denoising network are with clear objectives and functionalities, therefore, that leads to a model with good interpretability.

\begin{figure}[t]
    \centering
    \includegraphics[height=0.245\textwidth]{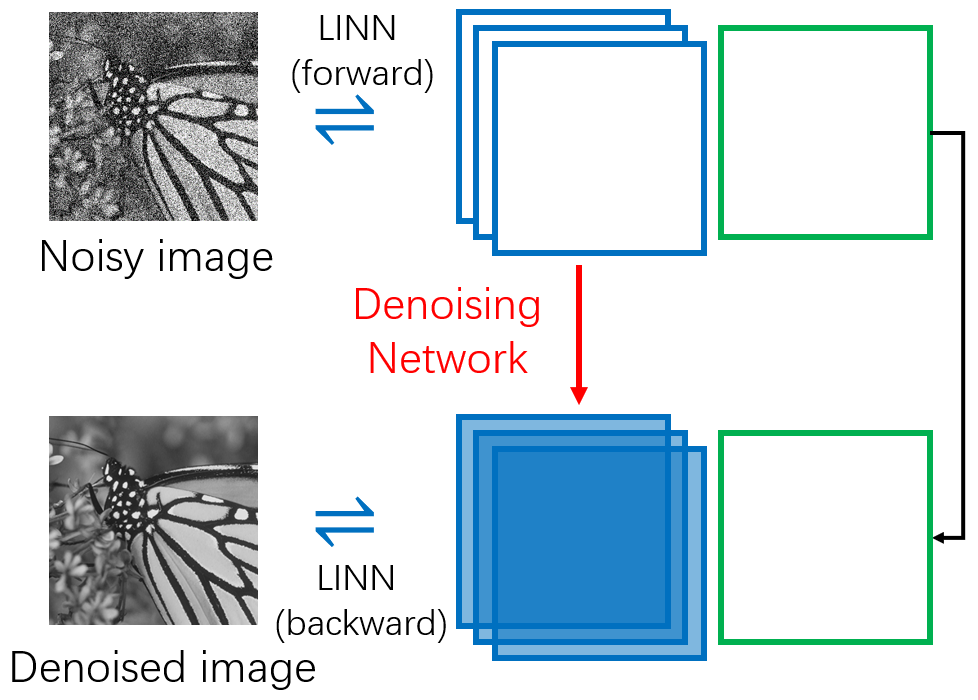}
    \caption{Overview of the proposed DnINN. The forward pass of the Lifting Inspired Invertible Neural Network (LINN) non-linearly transforms the input noisy image into coarse part (green) and detail part (blue). Denoising network will perform denoising operation on the detail part. The backward pass of the LINN will reconstruct the denoised image using the denoised detail part and the original coarse part.}
    \label{fig:DnINN_overview}
\end{figure}

DnINN method follows the principle of transform-based denoising \cite{donoho1994ideal} and has three main steps. First, the forward pass of the LINN performs a non-linear transform of the noisy image and results in 3 detail channels and 1 coarse channel. 
The detail channels should contain noise and high-frequency contents while the coarse channel should be less affected by noise. The decomposition can be iterated on the coarse channel to form a multi-scale architecture.
Then, the sparsity-inspired denoising network performs denoising on the detail channels, and we transfer the coarse channel to the reconstruction step. Finally, the denoised image is reconstructed by applying the backward pass of the LINN on the denoised representations. Figure \ref{fig:DnINN_overview} shows the overview of the proposed DnINN method. 

\subsection{Lifting Inspired Invertible Neural Network}\label{sec:inn}

The objective of LINN is to learn a transform in which the detail part contains noise and high-frequency contents and the coarse part would have less noise. Simple denoising operations would be possible to effectively remove the noise in the detail part.
Therefore, we would be able to recover a denoised image using the denoised detail part and the original coarse part.

\begin{figure}[t]
    \centering
    \includegraphics[height=0.245\textwidth]{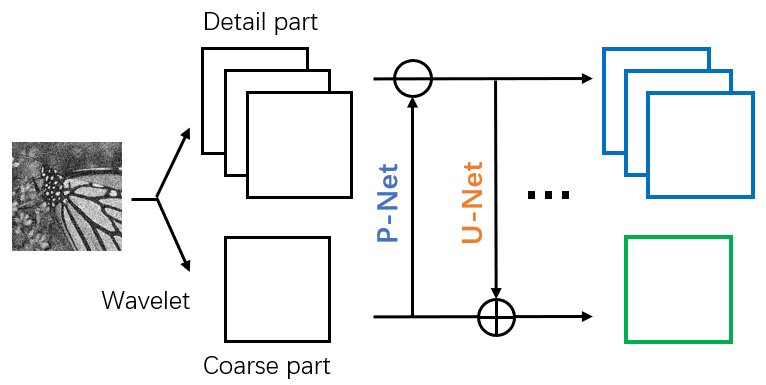}
    \caption{The forward pass of the Lifting Inspired Invertible Neural Network used in our DnINN. The wavelet transform is first used to transform the input image into a redundant representation which is splitted into the detail part and the coarse part. The predictor networks and the update networks sequentially update the detail part and the coarse part.}
    \label{fig:LINN}
\end{figure}

The neural network architecture used in DnINN is inspired by the lifting scheme \cite{sweldens1998lifting, daubechies1998factoring} and alternates prediction and update.
The LINN acts as a non-linear transform with perfect reconstruction condition. Compared to wavelet transforms and learned dictionaries, LINN can represent more complex features and is able to better separate signal and noise, and achieve better denoising performance. 
Figure \ref{fig:LINN} shows the schematic of the proposed LINN.

In the forward pass of LINN, the undecimated discrete wavelet transform (DWT) is applied on the input image $\bm{y} \in \mathbb{R}^{m \times n}$ to transform it into a redundant representation which is further splitted into detail part (the high-pass bands) $\bm{z}_d^{(0)} \in \mathbb{R}^{3 \times m \times n}$ and coarse part (the low-pass band) $\bm{z}_c^{(0)} \in \mathbb{R}^{1 \times m \times n}$. A Predictor network (P-Net) conditioned on $\bm{z}_c$ aims to predict the detail part to make the resultant detail part contain unique high-frequency contents. The Update network (U-Net) conditioned on $\bm{z}_d$ is used to adjust the coarse part to make it a better coarse version of the input image. There are $I$ pairs of P-Net and U-Net to sequentially update the detail part and the coarse part:
\begin{align} 
    \bm{z}_d^{(i)} &=  \bm{z}_d^{(i-1)} - P_i(\bm{z}_c^{(i-1)}), \\ 
    \bm{z}_c^{(i)} &=  \bm{z}_c^{(i-1)} + U_i(\bm{z}_d^{(i)}),
\end{align}
where $\bm{z}_d^{(i)}$ and $\bm{z}_c^{(i)}$ denotes the updated detail part and coarse part using the $i$-th P-Net $P_i(\cdot)$ and U-Net $U_i(\cdot)$, respectively.

In the backward pass of LINN, $\bm{z}_d^{(i-1)}$ and $\bm{z}_c^{(i-1)}$ can be recovered based on $\bm{z}_d^{(i)}$ and $\bm{z}_c^{(i)}$ and $P_i(\cdot)$ and $U_i(\cdot)$:
\begin{align} 
    \bm{z}_c^{(i-1)} &=  \bm{z}_c^{(i)} - U_i(\bm{z}_d^{(i)}),\\
    \bm{z}_d^{(i-1)} &=  \bm{z}_d^{(i)} + P_i(\bm{z}_c^{(i-1)}).
\end{align}

With $\bm{z}_d^{(0)}$ and $\bm{z}_c^{(0)}$, the target image can be recovered using inverse undecimated discrete wavelet transform (IDWT). If no operation is applied on $\bm{z}_d^{(I)}$ and $\bm{z}_c^{(I)}$ in the forward pass, the reconstructed target image would be identical to the input image, while if denoising operation is applied on $\bm{z}_c^{(I)}$, the reconstructed image would be a denoised image.

The P-Net and U-Net are constructed using convolution layers and soft-thresholding operators\footnote{Soft-thresholding is defined as $\mathcal{S}_{\lambda}(a) = \text{sgn}(a)\max(|a|-\lambda,0).$} which have been shown to be more effective than ReLU operator for some image processing tasks \cite{TSPDeepAM}.
Both P-Net and U-Net consist of $K$ layers of convolution and soft-thresholding operator followed by a convolution layer. We denote with $f \times f$ the spatial size of the convolution kernels and $M$ with the number of convolutional kernels.

\subsection{Denoising Networks}\label{sec:dnn}

The denoising network aims to denoise the detail part $\bm{z}_d^{I}$ by removing the noise features while retaining the image content. We model the denoising process as a $l_1$-minimization problem:
\begin{equation}
    \bm{g} = \arg \underset{\bm{g}}{\min} \Vert \bm{z}_d^I - \bm{g} \Vert_{2}^{2} + \lambda \Vert \bm{g} \Vert_{1},
    \label{eq:l1}
\end{equation}
where $\lambda$ is the regularization parameter.

\subsubsection{Soft-Thresholding Denoising Network}
Eqn. (\ref{eq:l1}) can be solved with closed-form solution using soft-thresholding operator. We model the denoising network as a single layer of soft-thresholding operators:
\begin{equation}
    \bm{g} = \mathcal{S}_{\bm{\lambda}}(\bm{z}_d^I),
\end{equation}
where $\bm{\lambda} \in \mathbb{R}^3$ is the soft-threshold vector to be learned.

Since the soft-thresholds $\bm{\lambda}$ relates to the noise level, we can adjust the learned soft-thresholds to adapt the learned denoising network to different noise levels. 

\subsubsection{LISTA Denoising Network}
To enrich the representation ability of the denoising network, we further over-parameterize Eqn. (\ref{eq:l1}) as:
\begin{equation}
    \bm{g} = \arg \underset{\bm{g}}{\min} \Vert \bm{z}_d^I - \bm{D}*\bm{g} \Vert_{2}^{2} + \lambda \Vert \bm{g} \Vert_{1},
    \label{eq:l1-over}
\end{equation}
where $\bm{D}$ denotes the convolution kernels and $*$ is the convolution operator.

The above $l_1$-minimization problem can be solved using Iterative Shrinkage-Thresholding Algorithm (ISTA) \cite{daubechies2004iterative}:
\begin{equation}
    \bm{g}_{t+1} = \mathcal{S}_{\lambda/\mu}\left( \left(\mathrm{I} - \frac{1}{\mu}\bm{D}^T * \bm{D} \right) * \bm{g}_t + \frac{1}{\mu} \bm{D}^T * \bm{z}_d^I \right),
\end{equation}
where $\mu$ is the step size.

\begin{figure}[t]
    \centering
	\subfigure[The soft-thresholding denoising network. $\bm{g}$ is the denoised image.]{
		\includegraphics[width=0.5\textwidth]{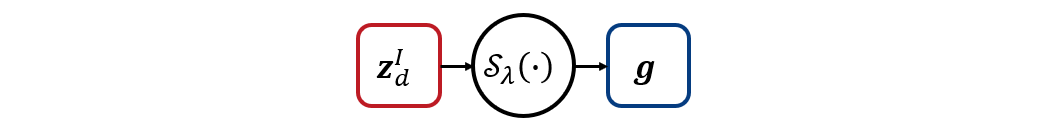}
		\label{fig:SoftNet}
		}
	\hspace*{\fill}	
	
	\centering
	\subfigure[The LISTA denoising network. $\bm{g}_T$ is the denoised image.]{
        \includegraphics[width=0.5\textwidth]{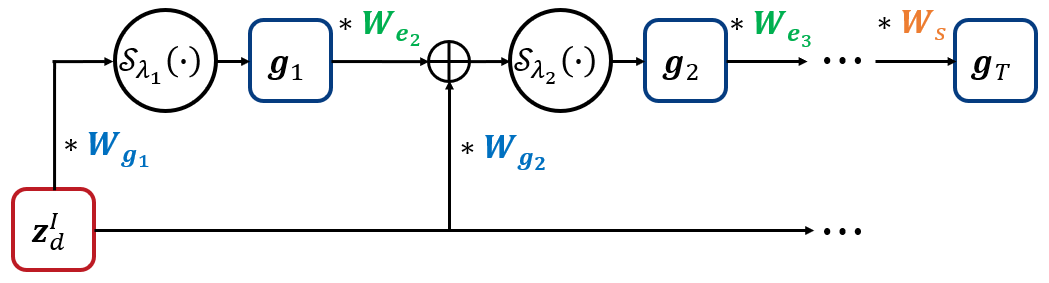}
        \label{fig:LISTA}
        }
        
    \caption{The proposed two denoising networks.}
    \label{fig:denoisingNet}
\end{figure}

We apply a learned version of ISTA (LISTA) \cite{gregor2010learning} as our second kind of denoising network. The schematic of the LISTA denoising network is shown in Figure \ref{fig:LISTA}.
There are $T$ layers of soft-thresholding operations in LISTA. For the $t$-th layer, there is a soft-thresholding operator: 
\begin{equation}
    \bm{g}_{t+1} = \mathcal{S}_{\bm{\lambda}_t}\left( \bm{W}_{et} * \bm{g}_t + \bm{W}_{gt} * \bm{z}_d^I \right),
\end{equation}
where $\bm{W}_{\bm{e}t}$ and $\bm{W}_{\bm{g}t}$ are two convolution kernels, and $\bm{\lambda}_t$ is the soft-threshold vector.

At the last layer of the LISTA denoising network, we add a synthesis convolution layer $\bm{W}_s$.

\subsection{Training Details}\label{sec:train}

The entire DnINN is trained end-to-end. Backpropagation algorithm with gradient descent is used to optimize all the parameters. We use Adam optimizer \cite{Kingma2014} with learning rate $l=1 \times 10^{-3}$ and $\beta=(0.9, 0.999)$.

The input image is the noisy image with noise drawn from $\mathcal{N}(0, \sigma^2)$, and the ground-truth image is the noise-free image. 
The learning objective is mean squared error between the denoised image and the noise-free image:
\begin{equation}
    \mathcal{L} = \frac{1}{N} \sum_{j=1}^{N} \Vert \bm{x}_j - \hat{\bm{x}}_j \Vert_2^2,
\end{equation}
where $N$ is the mini-batch size, and $\bm{x}_j$ and $\hat{\bm{x}}_j$ are the $j$-th noise-free image and denoised image in the mini-batch.

\section{Simulation Results} \label{sec:results}

In this section, we will show the implementation details, evaluate the proposed DnINN method, and visualize the results of the learned lifting scheme neural network and the denoising network. The source code is available at {https://github.com/JunjieHuangICL/LINN}.

\begin{table*}[]
    \centering
    \begin{tabular}{|l| C{1.6cm}|C{1.6cm}|C{1.6cm}|C{1.6cm}|}
        \hline
        Methods                         & Model Size          & $\sigma_N=15$    & $\sigma_N=25$   & $\sigma_N=50$     \\ \hline \hline
        BM3D \cite{dabov2007image}      & -                   &  31.07         & 28.57         & 25.63           \\ \hline
        WNNM \cite{gu2014weighted}      & -                   &  31.37         & 28.83         & 25.87           \\ \hline
        EPLL \cite{zoran2011learning}   & -                   &  31.21         & 28.68         & 25.67           \\ \hline
        TNRD \cite{chen2016trainable}   & $26.6 \times 10^3$  &  31.42         & 28.92         & 25.97           \\ \hline
        DnCNN \cite{zhang2017beyond}    & $556.0 \times 10^3$ &  31.70         & 29.19         & 26.20           \\ \hline
        DnINN$_{\text{ST}}$      & $134.7 \times 10^3$ &  31.58         & 29.08         & 26.14                 \\ \hline
        DnINN$_{\text{LISTA}}$   & $135.2 \times 10^3$ &  31.59       & 29.09          & 26.14           \\ \hline
        DnINN$_{\text{ST}}$ (2-scale)       & $269.3 \times 10^3$ & 31.62          & 29.14          & 26.19                 \\ \hline
        DnINN$_{\text{LISTA}}$ (2-scale)    & $270.3 \times 10^3$ & 31.63          & 29.14         & 26.20                 \\ \hline
    \end{tabular}
    \caption{The model size and PSNR (dB) results of different methods on BSD68 dataset on noise level $\sigma_N=15,25,50$.}
    \label{tab:compare}
\end{table*}

\begin{figure*}[t]
    \centering
    \centering
	\hspace*{\fill}
	\subfigure[Input noisy image ($\sigma=50$).]{
		\includegraphics[height=0.19\textwidth]{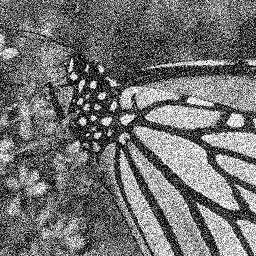}}
	\hfill
	\subfigure[$\bm{z}_d^I(1)$ before denoise.]{
		\includegraphics[height=0.19\textwidth]{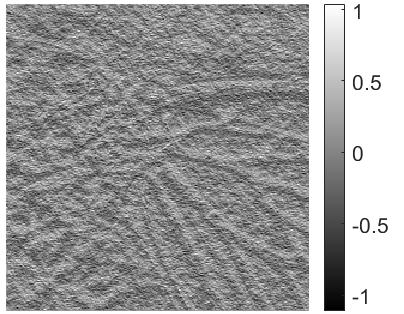}}
	\hfill
	\subfigure[$\bm{z}_d^I(2)$ before denoise.]{
		\includegraphics[height=0.19\textwidth]{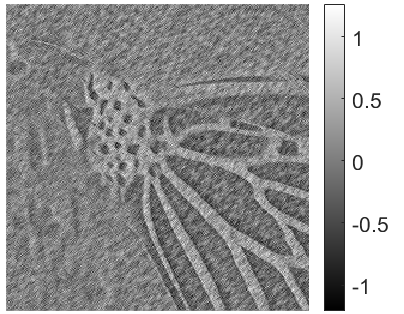}}
	\hfill
	\subfigure[$\bm{z}_d^I(3)$ before denoise.]{
		\includegraphics[height=0.19\textwidth]{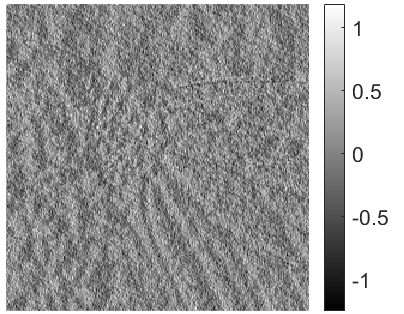}}
	\hspace*{\fill}
	
	\hspace*{\fill}
	\subfigure[The denoised image (PSNR=26.75 dB).]{
		\includegraphics[height=0.19\textwidth]{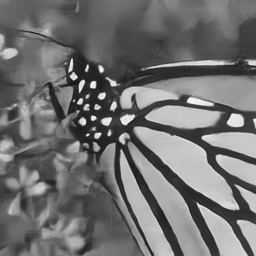}}
	\hfill
	\subfigure[$\bm{z}_d^I(1)$ after denoise.]{
		\includegraphics[height=0.19\textwidth]{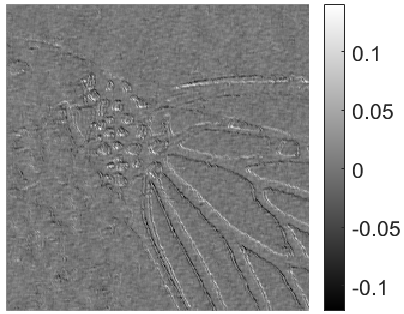}}
	\hfill
	\subfigure[$\bm{z}_d^I(2)$ after denoise.]{
		\includegraphics[height=0.19\textwidth]{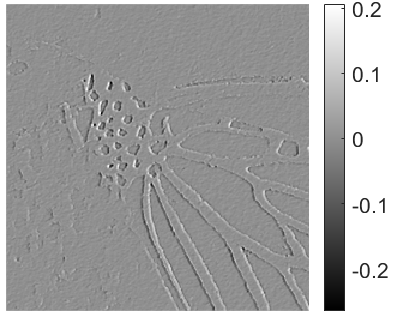}}
	\hfill
	\subfigure[$\bm{z}_d^I(3)$ after denoise.]{
		\includegraphics[height=0.19\textwidth]{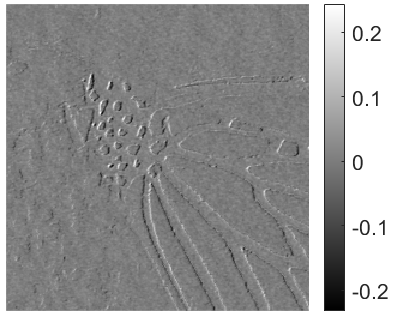}}
	\hspace*{\fill}
	
    \caption{Visualization of the input noisy image, the denoised image and the detail channels before and after denoising in DnINN$_{\text{LISTA}}$.
    ($\bm{z}_d^I(j)$ denotes the $j$-th channel of the detail part.)
    }
    \label{fig:visualize}
\end{figure*}

\subsection{Implementation Details}

We follow the training and testing settings in \cite{zhang2017beyond}. 400 training images of size $180 \times 180$ are used for training, the patch size is $40 \times 40$, and the number of patches for training is $9\times 10^4$. Three noise levels are considered $\sigma_N = 15, 25$ and 50. The testing images are the 68 natural images from the Berkeley segmentation dataset (BSD68) \cite{roth2009fields}.

The total number of training epochs is set to 30 and the learning rate decays from $1 \times 10^{-3}$ to $1 \times 10^{-4}$ at the 20-th epoch. The mini-batch size $N$ is set to 32. 

For the P-Net and U-Net, we set the number of layers as $K=8$, the spatial filter size as $f=3$, and the number of convolutional kernels at each layer as $M=16$. There are $I=4$ pairs of P-Net and U-Net in the neural network. We use the undecimated Haar wavelet transform in our implementation.
For LISTA denoising network, we set the number of layers $T=3$ and the spatial filter size as $3 \times 3$.

We denote the DnINN with soft-thresholding denoising network as DnINN$_{\text{ST}}$ and the DnINN with LISTA denoising network as DnINN$_{\text{LISTA}}$. For both cases, the number of parameters of the denoising network is very small.
In DnINN$_{\text{ST}}$, there are only 3 learnable soft-thresholds which controls the whole denoising operation. In DnINN$_{\text{LISTA}}$, the number of learnable parameter controlling the denoising operation is 495. 

\subsection{Comparison with Other Methods}

We compare the proposed DnINN method with classical image denoising methods: BM3D \cite{dabov2007image}, WNNM \cite{gu2014weighted}, EPLL \cite{zoran2011learning}, and DNN-based methods: TNRD \cite{chen2016trainable} and DnCNN \cite{zhang2017beyond}. 

Table \ref{tab:compare} shows the number of parameters and average PSNR (dB) results of different methods on BSD68 dataset on noise level $\sigma_N = 15, 25, 50$.
From Table \ref{tab:compare}, we can find that our proposed DnINN method achieves better performance compared to the classical image denoising methods and the TNRD \cite{chen2016trainable}. The performance of DnINN is comparable to that of DnCNN \cite{zhang2017beyond}, while DnINN only has $1/4$ learnable parameters as DnCNN.

We can see that the performance of DnINN$_{\text{LISTA}}$ is slightly better than that of DnINN$_{\text{ST}}$. This could be due to the stronger capability of LISTA to produce good sparse codes than a single layer of soft-thresholding operator.

We have further shown that multi-scale DnINN leads to improved performance. The first scale of DnINN removes most of the noise, and the second scale of DnINN further denoise the coarse part of the first scale and this leads to better denoising performance.

\subsection{Visualization of Learned Networks}

In Figure \ref{fig:visualize}, we show an exemplar image with visualization results. 
Figure \ref{fig:visualize} (a) and (e) show the input noisy image and the denoised image by DnINN$_{\text{LISTA}}$, respectively. The noise has been effectively removed.
Figure \ref{fig:visualize} (b)-(d) show the detail channels output from the lifting scheme neural network. They contain both noise and some image contents.
Figure \ref{fig:visualize} (f)-(h) show the corresponding detail channels after denoising. The noise has been significantly reduced. With the denoised detail channels (Figure \ref{fig:visualize} (f)-(h)) and the original coarse channel, the backward pass of the LINN reconstructs the denoised image shown in Figure \ref{fig:visualize} (e).

\begin{figure}
    \centering
    \includegraphics[width=0.48\textwidth]{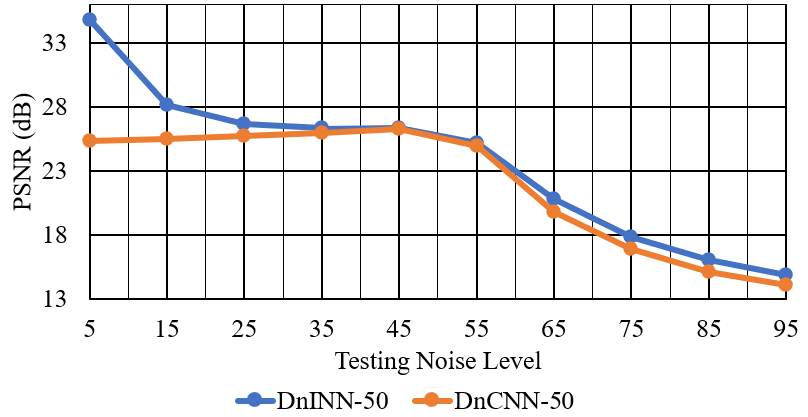}
    \caption{The performance of DnINN and DnCNN under different testing noise levels. Both DnINN and DnCNN are trained with $\sigma_N=50$. The testing noise levels are in the range of 5 and 95.}
    \label{fig:changeThres}
\end{figure}

\subsection{Adjust Soft-Thresholds in DnINN$_{\text{ST}}$}

The soft-thresholds in DnINN$_{\text{ST}}$ relates to the noise levels. We can therefore adjust the soft-thresholds according to testing noise level $\sigma_T$ to make the learned network adapt to the unseen noise level. The adjusted soft-thresholds are set to $\bm{\lambda} \times {\sigma_T^2}/{\sigma_N^2}$ ($\sigma_N$ is the noise level used to learn DnINN). 

Figure \ref{fig:changeThres} shows the performance of DnINN and DnCNN with $\sigma_N=50$ and $\sigma_T\in [5,95]$ with a step of 5. We can see that, by simply adjusting the soft-thresholds, DnINN can well adapt to unseen noise levels. 
When $\sigma_T > \sigma_N$, the performance of DnINN$_{\text{ST}}$ is consistently around 1 dB better than that of DnCNN. When $\sigma_T < \sigma_N$, DnINN$_{\text{ST}}$ achieves higher PSNR for images with smaller testing noise level, while at the same time the performance of DnCNN deteriorates for images with weaker noise.

\section{Conclusions} \label{sec:conclude}

In this paper, we proposed a image denoising invertible neural network (DnINN) method based on the principles of transform-based denoising. 
We propose an invertible lifting inspired invertible neural network to implement the non-linear transform with perfect reconstruction capability. Simple denoising networks can therefore easily remove the noise in the transform coefficients.
Simulation results show that the proposed DnINN method achieves comparable results as the DnCNN method while while using $1/4$ learnable parameters.

In the future, we plan to further improve the lifting scheme neural networks and apply DnINN to other image inverse problems.

\bibliographystyle{IEEEtran}
\bibliography{bibs}

\end{document}